\documentclass[12pt]{iopart}

\usepackage[english]{babel}
\usepackage{amsfonts}
\usepackage{amssymb}
\usepackage{amstext}
\usepackage{epsfig}

\newcommand{\beq}{\begin{equation}}
\newcommand{\eeq}{\end{equation}}
\newcommand{\barr}{\begin{eqnarray}}
\newcommand{\earr}{\end{eqnarray}}
\newcommand{\ket}[1]{\left\vert#1\right\rangle}

\newcommand{\fidav}{{\mathcal F}_{av}}
\newcommand{\purav}{{\mathcal P}_{av}}
\newcommand{\Ham}{\mathcal H}
\newcommand{\Id}{\mathbb I}
\newcommand{\eps}{\varepsilon}

\begin{document}
\title{Spin chain model for correlated quantum channels}

\author{Davide Rossini$^1$, Vittorio Giovannetti$^2$ and Simone Montangero$^2$}

\address{$^1$ \, International School for Advanced Studies SISSA/ISAS, 
  via Beirut 2-4, I-34014 Trieste, Italy}
\address{$^2$ \, NEST-CNR-INFM \& Scuola Normale Superiore,
  Piazza dei Cavalieri 7 , I-56126 Pisa, Italy \footnote[2]{URL: www.qti.sns.it}}

\date{\today}

\begin{abstract}

We analyze the quality of the quantum information transmission 
along a correlated quantum channel by studying the average fidelity 
between input and output states and the average output purity,
giving bounds for the entropy of the channel.
Noise correlations in the channel are modeled by the coupling of each channel
use with an element of a one dimensional interacting quantum spin chain.
Criticality of the environment chain is seen to emerge in the changes of the
fidelity and of the purity.

\end{abstract}


\section{Introduction}

The common scenario in quantum communication protocols is constituted by
two distant parties, Alice and Bob, who want to exchange information
through a quantum communication link.
Due to unavoidable noise in the channel, this cannot be perfectly
accomplished, and some strategies aimed to reduce communication errors
have to be employed. These are based on complex encoding/decoding operations
and on suitably tailoring the physical system that acts as a channel.
In this context the effect of noise on the quantum communication is typically
quantified by the so called {\it capacities} of the channel, that is the optimal
rates at which (quantum or classical) information can be reliably transmitted
in the limit of infinite channel uses~\cite{bennett98}.
The vast majority of the results obtained so far focused on the case of 
{\em memoryless} quantum channels, where 
the noise acts independently for each channel use. 
However, in real physical situations, correlations in the noise acting between
successive uses can be established. When this happens 
the communication line is said to be a {\em memory channel}, or more precisely, 
a {\em correlated channel}.
The analysis of these setups is much more demanding than the memoryless case,
and, at present, only a restricted class of them has been
solved~\cite{macchiav,bowen04,vgmancini05,vg05,daems07,kretsch05,darrigo07,caruso}.

\begin{figure}
  \begin{center}
    \includegraphics[width=10cm]{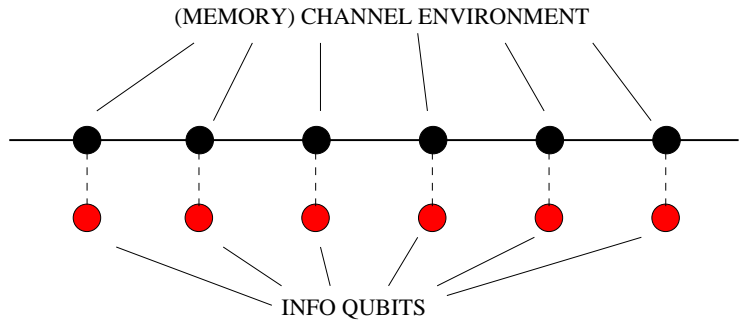}
    \caption{Model for memory channels: The red dots represent the
      channel uses (ordered, for instance, starting from the left to the
      right).
      The channel noise is modeled as a collection of local interactions
      between the channel uses and the many-body environment $E$
      (black dots).}
    \label{fig:model1}
  \end{center}
\end{figure}

Recently, a physical model for representing correlated channels has
been proposed in Refs.~\cite{vgmancini05,plenio07}, which, in the
context of Bosonic channels and qubit channels respectively, has established  
a direct connection between these systems and many-body physics. 
The setup discussed in these proposals  is depicted in Fig.~\ref{fig:model1}. 
Here Alice sends her messages to Bob by  encoding them into a $n$-long
sequence of information carriers $S$ (the red dots of the figure),
which model subsequent channel uses associated with $n$ independent
Bosonic modes~\cite{vgmancini05} or $n$ independent spins~\cite{plenio07}.
The correlated  noise of the channel is then described 
by assuming that each carrier interacts independently
with a corresponding element of a $n$-party  environment $E$ (sketched
with the connected  black dots of the figure), which,
in Refs.~\cite{vgmancini05} and~\cite{plenio07}, represents a multi-mode
Gaussian state and a many-body spin state, respectively.
Thus given an input state $\rho_S$ of the $n$ carriers,
the corresponding output state associated with the channel is 
\beq
  {\cal E}_{n} (\rho_S) = \mbox{Tr}_{E} [ {\cal U} \, (\rho_S \otimes \sigma_E)
    \, {\cal U}^\dag ] \;,
  \label{eq:SuperOp}
\eeq
where $\sigma_E$ is the joint input state of $E$ 
and the partial trace is performed over the environment.
In this equation ${\cal U}$ represents the unitary coupling between $S$ and $E$,
which is expressed as
\begin{eqnarray} \label{unitary}
{\cal U} = \bigotimes_{\ell=1}^n U^{(\ell)} \;,
\end{eqnarray}
with $U^{(\ell)}$ being the interaction between the $\ell$-th carrier and its
environmental counterpart (in Ref.~\cite{vgmancini05} these were beam-splitter
couplings, while in Ref.~\cite{plenio07} they were phase-gate couplings).
Within this framework, memoryless channels ${\cal E}_n = {\cal E}^{\otimes n}$
are obtained for factorizable environmental input states, while correlated
noise models correspond to correlated environmental states $\sigma_E$. 
Interestingly enough, in Ref.~\cite{plenio07} it was shown that it is possible
to relate the quantum capacity~\cite{qcap} of some specific
channels~(\ref{eq:SuperOp}) to the properties of the many-body environment $E$.

%

\begin{figure}
  \begin{center}
    \includegraphics[width=10cm]{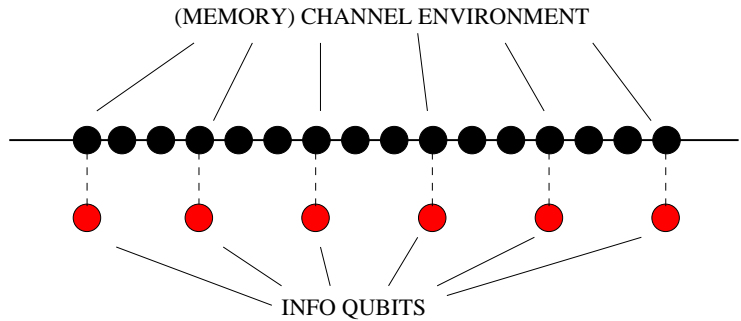}
    \caption{Generalized model of spin chain memory channels.
      As an example, in this figure we set $m=2$.}
    \label{fig:model2}
  \end{center}
\end{figure}

In this paper we discuss a variation of the model~(\ref{eq:SuperOp}),
which allows us to
adapt some of the techniques used in Ref.~\cite{rossini07} for
characterizing the decoherence effects induced by spin quantum baths,
in order to analyze the efficiency of a class of correlated qubits channels.
To do so we  consider a unitary coupling ${\cal U}$ 
that does not factorize as in Eq.~(\ref{unitary}).
Instead we assume $E$ to be a spin chain characterized by a free Hamiltonian
$\Ham_{E}$, whose elements interact with the carriers $S$ through the local
Hamiltonian $\Ham_{\rm int}$.
With this choice we write 
\begin{eqnarray}
{\cal U} = \exp[ -i \, (\Ham_E + \Ham_{\rm int}) \, t] \label{spinham}\;,
\end{eqnarray}
with the interaction time $t$ being a free parameter of the model.
In particular, as a chain Hamiltonian, we consider a spin-$1/2$ $XY$ model in a
transverse field, which can exhibit, in some parameters region, ground state
critical properties that greatly enhance spin correlations~\cite{sachdev00}.
Therefore the distance of the chain from criticality is non trivially
related to memory effects in the channel.
In the second part of the paper, we generalize the previous scheme by
introducing a given number of $m$ extra spins between any two
consecutive qubits, as shown in Fig.~\ref{fig:model2}.
In this case, we can use the number $m$ to modulate the memory effects.
 
\section{The Model} \label{sec:model}

As the environment $E$ of the system in Fig.~\ref{fig:model1}
we consider an interacting one-dimensional quantum spin-$1/2$ chain
described by an $XY$ exchange Hamiltonian in a transverse magnetic field:
\beq
  \Ham_E = - \frac{J}{2} \sum_{j=1}^{n} \left[ (1+\gamma) \sigma_j^x \sigma_{j+1}^{x}
  + (1 -\gamma) \sigma_j^y \sigma_{j+1}^{y} + 2 \lambda \sigma_j^z \right]
  \label{eq:HamEnv} \;,
\eeq
where $\sigma^\alpha_j$ (with $\alpha = x,y,z$) are the Pauli matrices of the
$j$-th spin, $J$ is the coupling strength between neighboring spins,
and $\lambda$ is the external field strength~\footnote{
Hereafter we always use open boundary conditions, therefore
we assume $\sigma^\alpha_n \sigma^\alpha_1 = 0$.}.
The model in Eq.~(\ref{eq:HamEnv}) for $0 < \gamma \leq 1$ belongs to
the Ising universality class, and has a critical point at $\lambda_c = 1$;
for $\gamma = 0$ it reduces to the $XX$ universality class, that is critical
for $|\lambda| \leq 1$~\cite{sachdev00}.

Following Ref.~\cite{rossini07}, we then assume that each carrier qubit is coupled
to one environmental spin element through the coupling Hamiltonian
\beq
\Ham_{\rm int}(j) = - \eps |e\rangle_j \langle e| \otimes \sigma^z_j \; ,
\eeq
where $\ket{g}_j$ and $\ket{e}_j$ respectively represent the ground and the
excited state of the $j$-th qubit.
Hence
the total
Hamiltonian $\Ham \equiv \Ham_E + \Ham_{\rm int}$ is given by
\beq
 \Ham = - \frac{J}{2} \sum_{j=1}^{n} \left[ (1+\gamma) \sigma_j^x \sigma_{j+1}^{x}
  + (1 -\gamma) \sigma_j^y \sigma_{j+1}^{y} + 2 \lambda \sigma_j^z \right]
  - \eps \sum_{j=1}^{n} |e\rangle_j \langle e| \sigma_j^z .
  \label{eq:Hamilt}
\eeq
Finally, as in Refs.~\cite{plenio07,rossini07}, 
we suppose that at time $t=0$ the
environment chain is prepared in the ground state $\ket{\varphi}_E$ of ${\cal H}_E$.
We then consider a generic input state $\ket{\psi}_S$ of the $n$ qubit 
carriers of the system (i.e.,
the input state of the red dots in Fig.~\ref{fig:model1}), and write it in the
computational basis:
\beq
  \ket{\psi}_S = \sum_{x} \alpha_x \ket{x}_S \, ,
  \label{eq:input}
\eeq
where $\alpha_x$ are complex probability amplitudes and the sum runs over
$N = 2^n$ possible choices of $x$, each of them being
a binary string of $n$ elements in which the $j$-th element 
is represented as $g$ or $e$, according to the state (ground or excited,
respectively) of the corresponding $j$-th qubit.

For each vector $\ket{x}_S$ we define ${\cal S}_x$ as the set of the
corresponding excited qubits (for instance, given $n = 5$ and
$\ket{x}_S = \ket{egeeg}_S$, then ${\cal S}_x$ contains the
$1^{\rm st}$, $3^{\rm rd}$ and $4^{\rm th}$ qubits).
After a time $t$, the global state of the qubits and the chain will then evolve
into
\beq
  \ket{\psi}_S \otimes \ket{\varphi}_E \quad \stackrel{\cal U}{\longrightarrow}
  \quad \sum_{x} \alpha_x \ket{x}_S \otimes {\cal U}_x \ket{\varphi}_E  \;,
\eeq
where ${\cal U}$ is the global evolution operator of Eq.~(\ref{spinham}),
while ${\cal U}_x \equiv  \exp[-i \, \Ham_E^{x} \, t]$ is associated
to the following {\em chain} Hamiltonian:
\beq
  \Ham_E^{x} \equiv \Ham_E - \eps \sum_{j\in {\cal S}_x} \sigma_j^z \, .
  \label{eq:HamChain}
\eeq
According to Eq.~(\ref{eq:SuperOp}), the channel output state is then described
by the density matrix
\beq
{\cal E}_n(|\psi\rangle_S\langle \psi|) =   \rho_S^\prime = \sum_{x,y} L_{xy} \; \alpha_x \alpha_y^* \; |x\rangle_S\langle y| 
  \label{eq:RhoOut} \;,
\eeq
where 
\beq
  L_{xy} \equiv {_E\langle} \varphi| \, {\cal U}_y^\dag \,
  {\cal U}_x \ket{\varphi}_E \;,
  \label{eq:GenEcho}
\eeq
can be seen as a generalized Loschmidt echo, denoting the scalar product
of the input environment state $\ket{\varphi}_E$ evolved with
${\cal U}_x$ and ${\cal U}_y$, respectively~\cite{gorin06}. 
These quantities
can be evaluated by first
mapping the Hamiltonian~(\ref{eq:HamChain}) into a free-fermion
model via a Jordan Wigner transformation~\cite{lieb61}
\beq
c_k = \exp \Bigg( i \pi \sum_{j=1}^{k-1} \sigma_j^+ \sigma_j^- \Bigg) \,
\sigma_k^-\;,
\eeq
where $\sigma^\pm = (\sigma^x \pm i \sigma^y)/2$, and then by
diagonalizing it with a Bogoliubov rotation of the Jordan Wigner
fermions $\{ c^\dagger_k , c_k \}_{k=1, \ldots, n}$.
This allows one to find an explicit expression of the Loschmidt echo
in terms of the determinant of a $2n \times 2n$ matrix
(see Ref.~\cite{rossini07} for details):
\beq
    L_{xy} = {_E\langle} \varphi \vert e^{i \Ham_E^x t} e^{-i
    \Ham_E^y t}  \vert \varphi \rangle_E =
  {\rm det} (\Id - \rho_0 + \rho_0 e^{i H_x t} e^{-i H_y t} ) \, ,
  \label{eq:det}
\eeq 
where $\Ham_k = \sum_{ij} [H_k]_{ij} \Psi^\dagger_i \Psi_j$,  
${\bf \Psi^\dagger} = \left( c_1^\dagger \ldots c_N^\dagger \,
c_1 \ldots c_N \right)$, and
$[\rho_0]_{ij} = {_E\langle} \varphi \vert \Psi_i^\dagger \Psi_j \vert
 \varphi \rangle_E$ are the two-point correlation functions of the chain.

\section{The channel}

The echoes~(\ref{eq:GenEcho}) provide a complete characterization of 
the correlated channel ${\cal E}_n$. In particular, since $L_{xx}=1$ for all $x$,
Eq.~(\ref{eq:RhoOut}) shows that the channel ${\cal E}_n$ is {\em unital},
i.e. it maps the completely mixed state $\frac{1}{N} \sum_x |x\rangle_S\langle x|$
into itself.
Furthermore, the $N\times N$ matrix of elements $L_{xy}/N$ coincides
with the {\em Choi-Jamiolkowski state}~\cite{BENGZY} of the map. The
latter is defined as the output density matrix obtained when
sending through the channel ${\cal E}_n$ half of the canonical 
maximally entangled state $|+\rangle_{SA} \equiv\frac{1}{\sqrt{N}}
\sum_x |x\rangle_S \otimes |x\rangle_A$ of the $N$-level system $S$,
i.e.  
\begin{eqnarray}
J({\cal E}_n) \equiv 
({\cal E}_n \otimes {\cal I}_A ) (|+\rangle_{SA}\langle +|) = 
 \sum_{x,y}\frac{L_{xy}}{N} \;  |x x \rangle_{SA} \langle yy | \label{choi}
\;,\end{eqnarray}
with $A$ being a $N$-dimensional ancillary system and  ${\cal I}_A$ being
the identity map. 
Similarly to the case analyzed in Ref.~\cite{plenio07}, this is a
{\em maximally correlated} state~\cite{rains01} whose 1-way distillable
entanglement is known to coincides with the ``hashing bound''~\cite{rains01,bennett99,devewin2005}:
\begin{eqnarray}
D_1(J({\cal E}_n)) = H(J_S({\cal E}_n)) - H(J({\cal E}_n)) = 
\log_2 N -  H(J({\cal E}_n))\;, \label{dist}
\end{eqnarray}
where $J_S({\cal E}_n) \equiv \mbox{Tr}_A [ J({\cal E}_n)]$ is the reduced
density matrix of $J({\cal E}_n)$ associated with the system $S$,
while $H(\cdot)= \mbox{Tr}[ (\cdot) \log_2 (\cdot) ]$ is the von
Neumann entropy.  
At least for the subclass of 
{\em forgetful} channels~\cite{kretsch05}, 
the regularized version of Eq.~(\ref{dist}) can then be used~\cite{bennett99,plenio07} 
to bound
the quantum capacity~\cite{bennett98,qcap}  of ${\cal E}_n$.  
This is~\footnote{The inequality~(\ref{distbound}) is a consequence of
  the fact that the quantum capacity $Q$ of a channel  
does not increase if we provide the communicating parties with a 1-way
(from the sender to the receiver) classical side communication
line~\cite{bennett99,vg05-1}. It is derived by constructing an explicit
quantum communication protocol in which {\em i)} Alice sends through
the channel half of the maximally entangled state $|+\rangle_{SA}$ to
Bob,  {\em ii)} the resulting state $J({\cal E}_n)$  is then 1-way
distilled obtaining $D_1(J({\cal E}_n))$ Bell pairs which, finally,
{\em iii)} are employed to teleport Alice messages to Bob.  It is
worth noticing that for the channel analyzed in Ref.~\cite{plenio07}
the right hand side of Eq.~(\ref{dist}) was also an upper bound for
$Q$.}  
\begin{eqnarray}
Q({\cal E}_n) \geqslant \lim_{n\rightarrow \infty} \frac{D_1(J_S({\cal E}_n))}{n} = 
1 -  \lim_{n\rightarrow \infty} \frac{H(J({\cal E}_n))}{n}\;. \label{distbound}
\end{eqnarray}
The quantity $H(J({\cal E}_n))$ corresponds to  
the {\em entropy of the channel} ${\cal E}_n$ of Ref.~\cite{roga08},
which can be used as an estimator of the channel noise.  
In our case it has also a simple interpretation in terms of the
properties of the many-body system $E$: it measures the entropy of the
ground state $|\varphi\rangle_E$ after it has evolved through a random
application of  
the perturbed unitaries ${\cal U}_x$~\footnote{
This is a trivial consequence of the fact that $J({\cal E}_n)$ is the reduced
density matrix of the pure state $ \frac{1}{\sqrt{N}}
\sum_x |xx\rangle_{SA}\otimes {\cal U}_x |\varphi\rangle_E$
tracing out the environment, and of the fact that the von Neumann entropies of
the reduced density matrices of a pure bipartite system coincide.},
i.e.
\begin{eqnarray}
H(J({\cal E}_n)) = H(\sigma_E^\prime)\;, \qquad \mbox{with} \qquad 
\sigma_E^\prime \equiv \frac{1}{N} \sum_x {\cal U}_x \;  
|\varphi\rangle_E\langle \varphi| \; {\cal U}^\dag_x
 \label{entE}\;.
\end{eqnarray}
Unfortunately, for large $n$ the computation of the von Neumann entropy of
the state $J({\cal E}_n)$ is  impractical both analytically and numerically,
since it requires to evaluate an exponential number of $L_{xy}$ elements.
Interestingly enough, however, we can simplify our analysis 
by considering the fidelity between $J({\cal E}_n)$ and its input counterpart
$|+\rangle_{SA}$ (see Eq.~(\ref{choi})). 
As discussed in the following section, this is a relevant information
theoretical quantity, since it is directly related to 
the {\em average fidelity} between input and output state of the
channel ${\cal E}_n$ and provides us an upper bound for $H(J({\cal E}_n))$.
Similarly we can compute the purity of $J({\cal E}_n)$ 
which, on one hand, gives a lower bound for $H(J({\cal E}_n))$,
while, on the other hand, it is directly related to
the average channel output {purity} of the map ${\cal E}_n$. 


\begin{figure}
  \begin{center}
    \includegraphics[width=13cm]{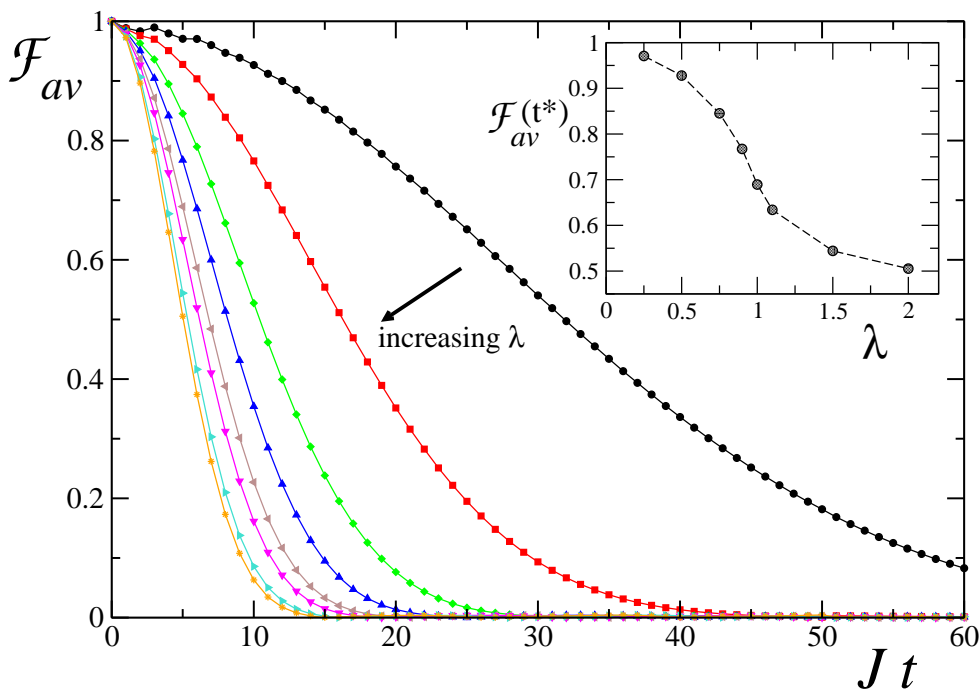}
    \caption{Averaged channel fidelity as a function of the interaction time,
      for different values of the transverse field $\lambda$: from right to left
      $\lambda = 0.25, \, 0.5, \, 0.75, \, 0.9, \, 1, \, 1.1, \, 1.5, \, 2$.
      Here we simulated a channel of $n=50$ qubits coupled to an Ising chain,
      and set an interaction strength $\eps = 0.05$; the fidelity has been
      evaluated by sampling over $N_{av} = 5 \times 10^4$ randomly chosen
      initial conditions.
      In the inset we plot the averaged fidelity at a fixed interaction time
      $J t^* = 5$, as a function of $\lambda$.}
    \label{fig:L50_Bvar}
  \end{center}
\end{figure}

\section{Average transmission fidelity} \label{sec:fidelity}

According to Eq.~(\ref{choi}), the fidelity between the Choi-Jamiolkowski state
$J({\cal E}_n)$ and its input counterpart 
coincides with the average value of the Loschmidt echoes $L_{xy}$, i.e.  
\begin{eqnarray}
{\cal F} \equiv {}_{SA}\langle + | J({\cal E}_n) | +\rangle_{SA}
= \frac{1}{N^2} \sum_{x,y} L_{xy}
\label{fido}\;.
\end{eqnarray}  
Even without computing all the $L_{xy}$, this quantity can be numerically
evaluated by performing a sampling over $N_{av}$ randomly chosen
couples ($x, y$) of initial conditions, and averaging over
them~\footnote{
We numerically checked the convergence of ${\cal  F}$ with $N_{av}$.
We first considered a situation with a few number of qubits ($n \leq 10$),
such to compare sampled averages, ${\cal F}_{av}$, with exact averages
over all possible events, ${\cal  F}_{ex}={\cal F}$.
We found that, already at $N_{av} = 10^4$, absolute differences 
$\vert {\cal F}_{av} - {\cal F }_{ex} \vert$
are always less than $2 \times 10^{-2}$, while at $N_{av} = 5 \times 10^4$
the error is less than $5 \times 10^{-3}$,
independently of the values of the interaction time $t$, the transverse field
$\lambda$ and the system size $n$.
In a second time, we simulated systems with definitely larger sizes
($n \approx 50$) and simply check the convergence of ${\cal F}_{av}$
with $N_{av}$. Differences between fidelities with $N_{av} = 10^4$ and
$N_{av} = 5 \times 10^4$ are of the same order as the deviation of the curve
with $N_{av} = 10^4$ from the exact one for small sizes.
Therefore we can reliably affirm that fidelity results with $N_{av} = 5 \times 10^4$
are exact, up to an absolute error of order $5 \times 10^{-3}$.
}:
\beq \label{effeap}
 {\cal F} \approx {\cal F}_{av} \equiv \frac{1}{N_{av}} \sum_{(x,y)=1}^{N_{av}} {\rm Re} [L_{xy}] \, ,
\eeq
(where we used the fact that $L_{yx} = L_{xy}^*$).
The quantity  ${\cal F}$  provides an upper bound for $H(J({\cal E}_n))$ through
the quantum Fano inequality~\cite{nielsen}, i.e. 
\begin{eqnarray}
H(J({\cal E}_n)) \leqslant H_2({\cal F}) + (1 -{\cal F}) \log_2 ( 4^n -1 ) \leqslant H_2({\cal F}) +  2 n (1 -{\cal F}) \;, 
\label{upper}
\end{eqnarray}
where $H_2(\cdot) = - (\cdot) \log_2 (\cdot) - [1 - (\cdot)] \log_2 [1- (\cdot)]$
is the binary entropy function\footnote{Equation~(\ref{upper}) can be
easily derived by noticing that $H(J({\cal E}_n))$ and ${\cal F}$ coincide,
respectively, with the {\em exchange entropy} and {\em entanglement fidelity}
of the channel ${\cal E}_n$ associated
with the maximally mixed state $\mathbb{I}_S/2^n$ of $S$.}.  
Furthermore ${\cal F}$ is directly related to the average
transmission fidelity $\langle F\rangle$ of the map ${\cal E}_n$. For
a given pure input state (\ref{eq:input}), the transmission fidelity is 
\beq
  F(\psi) \equiv {_S \langle \psi}|  {\cal E}_n(|\psi\rangle_S\langle \psi|)
 \ket{\psi}_S
  = \sum_{x,y} L_{xy} |\alpha_x|^2 |\alpha_y|^2  \; .
\eeq
Taking the average with respect to all possible inputs, we get
\beq
  \langle F\rangle  = \sum_{x,y} L_{xy} \, p_{xy} \, ,
  \label{eq:FidCh}
\eeq
where $p_{xy} = \langle |\alpha_x|^2 |\alpha_y|^2\rangle$ with
$\langle ... \rangle$
being the average with respect to the uniform Haar measure.
\begin{figure}[ht]
  \begin{center}
    \includegraphics[width=13cm]{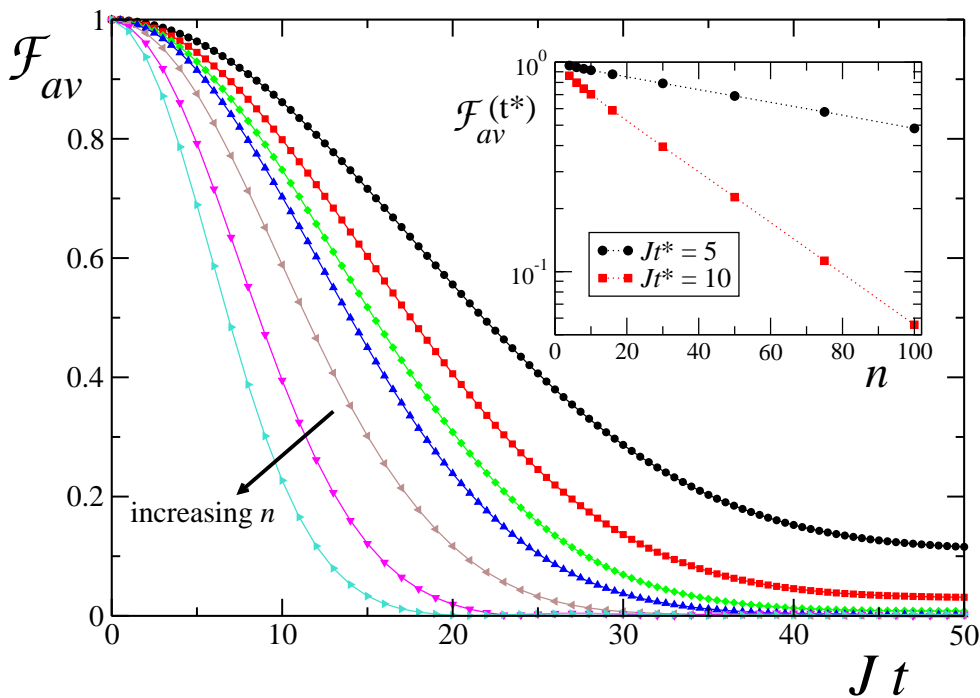}
    \caption{Fidelity for a channel coupled to an Ising chain with
      $\lambda=1$ and different qubit numbers $n$: from right to left
      $n= 4, \, 6, \, 8, \, 10, \, 16, \, 30, \, 50$;
      the interaction strength is kept fixed at $\eps = 0.05$;
      data are averaged over $N_{av} = 5 \times 10^4$ configurations.
      Inset: $\fidav$ at a fixed time $t^*$, as a function of $n$.}
    \label{fig:B100_Nvar}
  \end{center}
\end{figure}
The probability distribution $p_{xy}$ can be computed by using simple
geometrical arguments~\cite{lubkin78}. As shown in~\ref{appa}, this
yields $p_{xy} = \frac{1+2\, \delta_{x,y}}{N(N+2)}$ and hence: 
\beq \label{ave}
  \langle F \rangle = \left( \frac{2}{N(N+2)} \sum_{x > y} {\rm Re} [L_{xy}] \right)
  + \frac{3}{N+2} \, ,
\eeq
where we used the fact that $L_{xx} = 1$ and $L_{xy} = L_{yx}^*$.
Therefore, from Eq.~(\ref{fido}) we get
\begin{eqnarray}
\langle F \rangle = \frac{N}{N+2} \; {\cal F} + \frac{2}{N+2}
\label{fafa}\;.
\end{eqnarray}
The fidelities ${\cal F}$ and $\langle F \rangle$ are not directly related to
the channel quantum capacity, nonetheless, as in Eq.~(\ref{upper}), they can be used
to derive bounds  for $Q$ \footnote{In particular from Eq.~(\ref{upper}) and
Eq.~(\ref{distbound}) one gets
$Q \geqslant 1 - 2 \lim_{n\rightarrow \infty} {\cal F}$.}.
More generally, values near to unity of the fidelity between the output channel 
states and their corresponding input states, are indicative of a fairly noiseless
communication line.
On the contrary, values of the transmission fidelities close to zero, while
indicating output states nearly orthogonal to their input counterparts, 
do not necessarily imply null or low capacities, since such huge discrepancies
between inputs and outputs could still be corrected by a proper
encoding and decoding strategy
(e.g. consider the case of a channel which simply rotate the system states).

\begin{figure}[ht]
  \begin{center}
    \includegraphics[width=13cm]{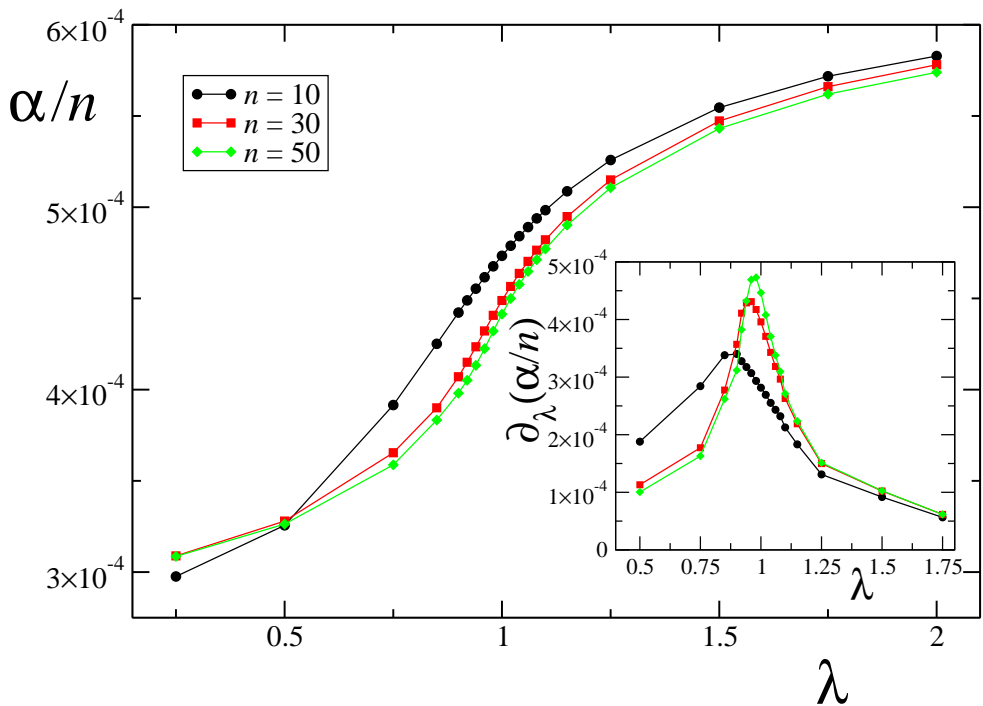}
    \caption{Short-time Gaussian decay rate $\alpha$ as a function of the
      transverse field $\lambda$. We rescaled $\alpha$ with the system size
      according to the scaling of Fig.~\ref{fig:B100_Nvar}.
      In the inset we plot the first derivative of the same curves in the main
      panel, with respect to $\lambda$.}
    \label{fig:Alpha_scal}
  \end{center}
\end{figure}

Equation~(\ref{eq:det}) allows us to numerically compute the averaged
transmission fidelity~(\ref{effeap}): numerical results 
in this and in the next sections are given for the case $\gamma=1$, 
i.e. we study the correlated quantum channel defined by the Ising model.
The behavior of the averaged channel fidelity $\fidav$, defined in Eq.~(\ref{effeap})
and related to the memory channel scheme of Fig.~\ref{fig:model1},
with respect to the interaction time $t$ (the free model parameter)
is shown in Fig.~\ref{fig:L50_Bvar}: the plots are given for $n=50$ qubits,
each of them coupled to one spin of an Ising chain with a coupling strength
$\eps=0.05$. Different curves stand for different values of the 
transverse magnetic field $\lambda$: as it can be clearly seen,
the fidelity $\fidav$ decays as a Gaussian in time, irrespective of the field
strength $\lambda$~\footnote{
For times longer than those in the scales of
Figs.~\ref{fig:L50_Bvar} and~\ref{fig:B100_Nvar}, 
revivals of the fidelity are present. See \ref{appB} for details.
}.
The signature of criticality in the environment chain can be identified
by studying the function 
$\fidav (\lambda, t)$ for fixed interaction time $t^*$: the inset of 
Fig.~\ref{fig:L50_Bvar} displays a non analytic behavior for the derivative 
of $\fidav (\lambda, t^*)$ with respect to $\lambda$ at the critical point 
$\lambda_c = 1$. This non analyticity will be clearer in the following,
where the size scaling will be considered.
In Fig.~\ref{fig:B100_Nvar} we study the behavior of $\fidav$ with the
number $n$ of qubits, at a fixed value of transverse magnetic field $\lambda = 1$.
As it is shown in the inset, at a given interaction time $t^*$,
the fidelity $\fidav (\lambda_c,t^*)$
decays exponentially with $n$  (the same behavior is found when 
$\lambda \ne 1$). This dependence of the decay rate implies that the 
average fidelity $\fidav$ can be fitted by:
\beq
\fidav \sim e^{- \alpha t^2} \qquad \textrm{ with } \quad \alpha \propto n \, ;
\eeq
in other words, the Gaussian decay rate is extensive.
Indeed, since the fidelity is a global quantity that 
describes the evolution of the state of the whole $n$-body system,
it should start decaying as a Gaussian (at least at small times)~\cite{peres},
with a reasonably extensive decay ratio.
This prediction is confirmed by the results of Fig.~\ref{fig:Alpha_scal},
where we report the decay rate as a function 
of the transverse magnetic field for different system sizes $n$. 
In proximity of the critical point, the decay rate undergoes a 
sudden change, which becomes more evident when increasing the system size.  
The signature of criticality at $\lambda_c=1$ and the finite size effects
can be better analyzed by looking at the derivative of the decay rate with respect
to the transverse field: The inset of Fig.~\ref{fig:Alpha_scal}
clearly show that $\partial_\lambda \alpha$ exhibits a non analytic behavior
at the critical point $\lambda_c$, at the thermodynamical limit.
Notice also that, due to finite size effects, the maximum of 
$\partial_\lambda \alpha$ does not coincide exactly with the critical point,
that can be rigorously defined only at the thermodynamical limit,
but occurs at a slightly smaller value of $\lambda$.
However, we checked that a finite size scaling gives the right prediction
of the critical point located at $\lambda = 1$.

\section{Average output purity}

Another quantity that can be evaluated with relatively little numerical
effort is the purity of the Choi-Jamiolkowski state $J({\cal E}_n)$, i.e.
\begin{eqnarray} \label{puri2}
{\cal P}_2 \equiv \mbox{Tr} [ J({\cal E}_n)^2 ] = \frac{1}{N^2} \sum_{xy} |L_{xy}|^2 \;.
\end{eqnarray}
As in the case of ${\cal F}$, this can be computed by approximating
the summation with a random sampling, i.e.  
\beq \label{puriap}
 {\cal P}_2 \approx {\cal P}_{av} \equiv \frac{1}{N_{av}} \sum_{(x,y)=1}^{N_{av}} |L_{xy}|^2\, .
\eeq
The quantity~(\ref{puri2}) provides us two important pieces of information.
First of all, it yields a useful bound on the channel entropy $H(J({\cal E}_n))$.
This follows from the inequality~\cite{renyi}
\begin{eqnarray}
H(J({\cal E}_n)) \geqslant H_2(J({\cal E}_n)) = - \log_2  {\cal P}_2 \;,
\end{eqnarray} 
with $H_2(\cdot) \equiv - \log_2 \mbox{Tr}[ (\cdot)^2]$ 
being the R\'{e}nyi entropy of order 2.
Furthermore ${\cal P}_2$ is directly related to the average 
output purity $\langle P_2 \rangle$ of the channel ${\cal E}_n$.
This is obtained by averaging over all possible inputs $|\psi\rangle_{S}$ the purity
of the output state ${\cal E}_n(|\psi\rangle_S\langle \psi|)$, i.e. 
\beq
\langle P_2 \rangle  \equiv \langle {\rm Tr} \left[ ({\cal E}_n(|\psi\rangle_S\langle \psi|)
)^2 \right] \rangle =
\sum_{x,y} \vert L_{xy} \vert^2 \, \langle \vert \alpha_x \vert^2 \vert \alpha_y \vert^2 \rangle =
\sum_{x,y} \vert L_{xy} \vert^2 \, p_{xy} \,,
\eeq
where we used 
Eq.~(\ref{eq:RhoOut}) and where $p_{xy}$ are the probabilities defined 
in Eq.~(\ref{eq:probab}).
According to Eq.~(\ref{puri2}) this yields,
\beq
\langle P_2 \rangle  = \frac{N}{N+2} {\cal P}_2 + \frac{2}{N+2} \label{fine}\;.
\eeq
\begin{figure}
  \begin{center}
    \includegraphics[width=13cm]{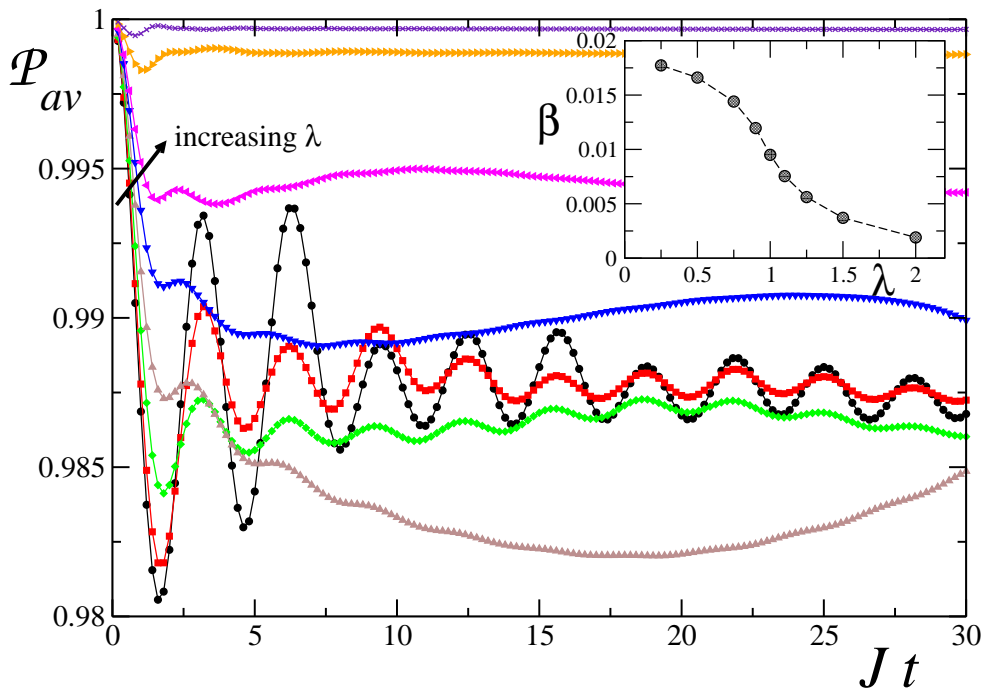}
    \caption{Averaged purity of the channel output state as a function
      of the interaction time, for various transverse field strengths:
      from left to right $\lambda = 0.25$ (black circles),
      0.5 (red squares), 0.75 (green diamonds), 0.9 (brown triangles up),
      1 (blue triangles down), 1.1 (magenta triangles left),
      1.5 (orange triangles right), 2 (violet crosses).
      Here we simulated a channel of $n=30$ qubits coupled to an Ising chain,
      set an interaction strength $\eps = 0.05$, and averaged over
      $N_{av} = 5 \times 10^4$ random initial conditions.
      In the inset we plot the short-time Gaussian decay rate $\beta$
      as a function of $\lambda$.}
    \label{fig:Purity_L30}
  \end{center}
\end{figure}
The average purity is a rather fair indicator of the noise induced by the coupling
to the environment: if the carrier qubits get strongly entangled with the environment,
$P_2$ is greatly reduced from the unit value (for large $n$ it will tend to zero);
on the other hand, a channel which simply unitarily rotates the carrier states
has a unit purity.
However, we should stress that also the purity may intrinsically fail
as a transmission quality quantifier: there are strongly noisy channels
with very high output purity (consider, for example, the channel which maps
each input state into the same pure output state).

Then, we study the average channel purity $\purav$ as a function of
the model free parameter, the interaction time $t$. The results are 
reported in Fig.~\ref{fig:Purity_L30} for a chain of $n=30$ qubits and different
values of the transverse field $\lambda$. We notice qualitatively different
behaviors depending on the values of the transverse field $\lambda$:
If $\lambda < \lambda_c$ the averaged purity oscillates in time and 
asymptotically tends to an average constant value; as far as the 
critical point is approached, $\purav$ drops to smaller values 
(revivals are again due to finite system size effects),  
reflecting the fact that at criticality correlations between the qubits 
and the environment are stronger.
Crossing the critical point, in the $z$-ordered phase 
($\lambda > \lambda_c$) the purity is generally higher and 
asymptotically takes values very close to the unit value. 
This can be easily understood in the limit $\lambda \to \infty$: in this case 
the spins in the chain are ``freezed'' along the field direction and they cannot 
couple with anything else, resulting in a watchdog-like effect~\cite{peres}. 
Independently of the transverse field value, the average purity $\purav$ 
decays as a Gaussian $\purav \sim e^{- \beta t^2}$  in the short time limit. 
As for the averaged fidelity $\fidav$, we have then analyzed the decay
rate $\beta$ as a function of $\lambda$: as before, $\beta$ exhibits
a signature of criticality via a divergence, at the thermodynamical limit, 
in its first derivative with respect to $\lambda$ 
(see the inset of Fig.~\ref{fig:Purity_L30}).
\begin{figure}
  \begin{center}
    \includegraphics[width=13cm]{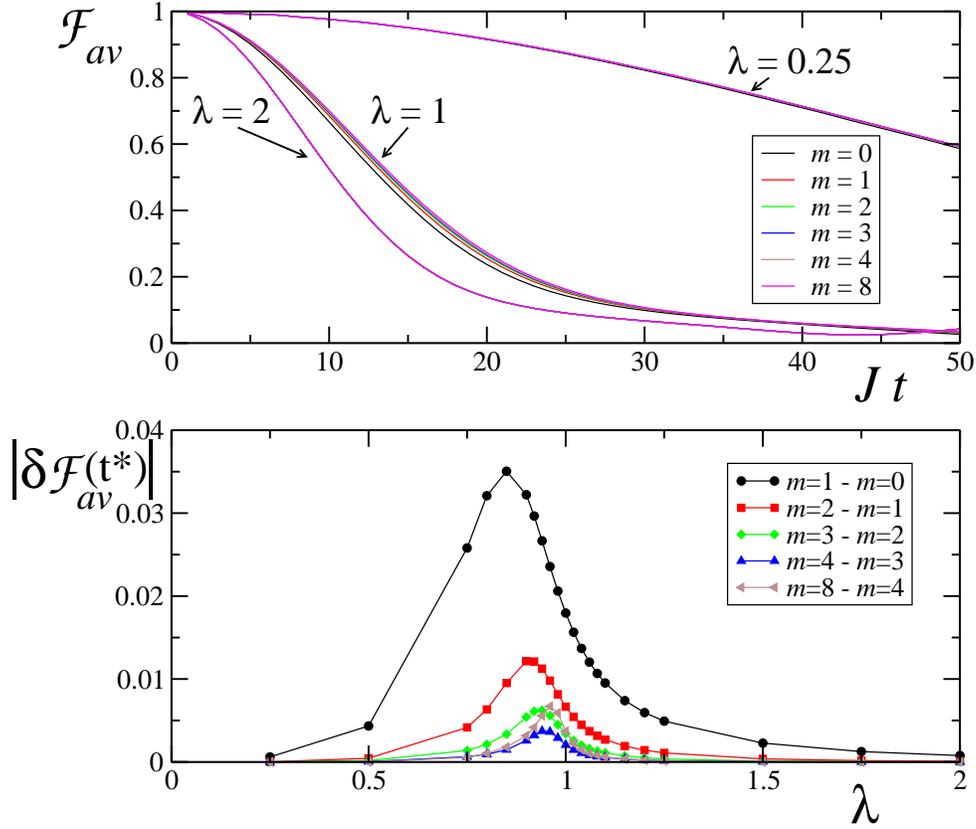}
    \caption{Upper panel: average channel fidelity for the generalized model
      in Fig.~\ref{fig:model2}, with $n=12$ qubits, $\eps = 0.05$
      (averages have been performed over $N_{av} = 10^4$ initial states).
      The various curves are for different numbers $m$ of spins between
      two consecutive qubits, and different values of transverse magnetic
      fields $\lambda = 0.25, \, 1, \, 2$.
      Lower panel: absolute differences in the fidelities between
      configurations at various $m$, as a function of $\lambda$ and at a
      fixed interaction time $J t^* = 10$.}
    \label{fig:Scheme2_L12}
  \end{center}
\end{figure}

\begin{figure}
  \begin{center}
    \includegraphics[width=13cm]{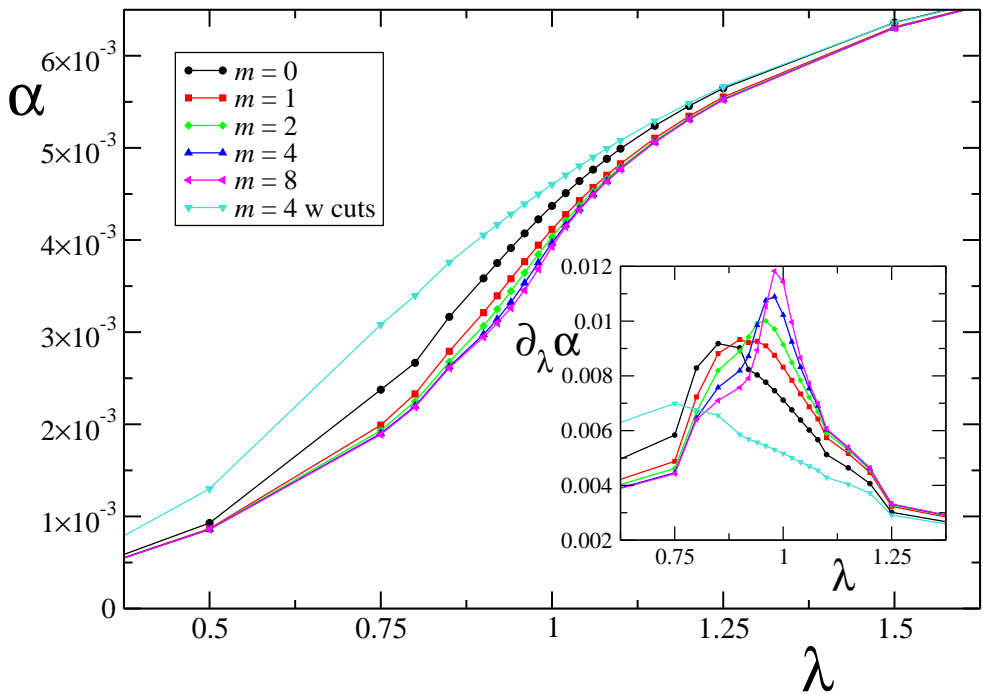}
    \caption{Gaussian decay rate of the averaged fidelity for model
      in Fig.~\ref{fig:model2}, as a function of the transverse field $\lambda$.
      The various curves stand for different numbers of spins $m$ between
      two consecutive qubits (here $n=12$, $\eps = 0.05$, $N_{av} = 10^4$).
      In the inset we show the first derivative with respect to $\lambda$
      of the curves in the main panel.}
    \label{fig:Scheme2_L12_Alpha}
  \end{center}
\end{figure}

\section{Generalized model}

We finally concentrate on the generalized model depicted in 
Fig.~\ref{fig:model2}, where a certain number $m$ of environment 
spins are present between two consecutive spins coupled to the qubits. 
The richness of the model, that is characterized by a large number of parameters,
and by a global size which grows both with $n$ and $m$,
requires a huge numerical effort in order to simulate it,
therefore we decided to analyze only the average channel fidelity $\fidav$.
In the upper panel of Fig.~\ref{fig:Scheme2_L12} we show $\fidav$
as a function of the interaction time $t$ for different values of $m$
and three values of the transverse field $\lambda = 0.25, \, 1 , \, 2$;
we fix a number of qubits $n = 12$ and an interaction strength $\eps = 0.05$.
Hereafter we will concentrate on this case as a typical result,
as we performed some checks with larger numbers 
of channel uses ($n=30, \, 50$ and $m=0, \, 1, \, 2$), and found 
qualitatively analogous results.
We immediately observe that differences for various $m$ are tiny,
even if the fidelity generally tends to increase when increasing $m$;
the sensibility with $m$ suddenly enhances at criticality ($\lambda_c = 1$),
where correlations in the environment decay much slower than in the other cases.
On the other hand, when $\lambda$ is far from $\lambda_c$, 
differences between fidelities upon a variation of $m$ are greatly suppressed,
and the generalized model mostly behaves as the model in Fig.~\ref{fig:model1}.
Again, this reflects the fact that, out of criticality, each qubit is mostly
influenced only by the spin that is coupled to, as the spin does not 
exchange correlations with the other environmental spins.
The resulting channel properties are then defined only by the 
local properties of the chain. On the contrary, at criticality, the spins
are correlated and then the resulting channel properties are influenced 
by the distance of the spins coupled with the qubits.
In the lower panel of Fig.~\ref{fig:Scheme2_L12} we explicitly plot
the differences in the fidelities for various $m$ as a function of $\lambda$,
and for a fixed interaction time; a peak in proximity of $\lambda_c$ is
clearly visible.
We point out that, as already noted at the end of Sec.~\ref{sec:fidelity}
concerning the size scaling of the fidelity, the maximum in the differences
does not occur exactly at the critical point.

The sensitivity to criticality is  again demonstrated by the averaged
fidelity Gaussian decay rate $\alpha$ as a function of $\lambda$, as shown
in Fig.~\ref{fig:Scheme2_L12_Alpha} for different values of $m$:
the first derivative in the inset has a maximum in correspondence of a value
that approaches the critical point $\lambda_c$ at the thermodynamical limit.
Indeed, increasing $m$ is equivalent to approaching the thermodynamical limit
of the chain, thus resulting in an increase of the quantum phase transition effects.
A double check of this comes from the cyan triangles-down curve
of Fig.~\ref{fig:Scheme2_L12_Alpha}: in this case we 
take $m=4$, but we break one of the links between two intermediate spins.
The environment is then formed by disconnected chains, each of them made
up by $5$ spins, therefore the system cannot undergo a phase transition
in the limit $n \to \infty$: the signature of criticality has completely
disappeared.

\section{Conclusions}

In conclusion we have introduced and characterized a class of 
correlated quantum channels, and we have given bounds for its entropy
by means of the averaged channel fidelity $\fidav$ and purity $\purav$. 
Even though in general these bounds might not be strict, we give a
characterization of the channel in terms of quantities that have a clear
meaning from the point of view of the many body model we have introduced.

In the case of an environment defined by a quantum Ising chain, we have shown
that the averaged channel purity and the fidelity depend on the environment
parameters and are strongly influenced by spin correlations inside it, in
particular by the fact whether the environment is critical or not. 
We expect that some different environment models, such as, for example, the $XY$
spin chain, will behave qualitatively similarly of what found in this work,
as it belongs to the same universality class.
This might not be the case for other models, like the Heisenberg chain,
which will be object of further study in the near future.

\section*{Acknowledgments}

We thank R.~Fazio for discussions and support, F. Caruso for comments, 
and D.~Burgarth for pointing out Ref.~\cite{roga08}.
This work have been supported by the ``Quantum Information Program'' 
of Centro De Giorgi of Scuola Normale Superiore.

\appendix 

\section{}\label{appa}

The probability $p_{xy}$ of Eq.~(\ref{eq:FidCh}) can be computed as follows:
we first define $r_x \equiv \vert \alpha_x \vert$ and convert the string $x$ 
into a decimal number from $1$ to $N = 2^n$ 
by trivially identifying $g \equiv 0$ and $e \equiv 1$.
The average over a uniform distribution of all pure input states on the Bloch
hypersphere for $n$ qubits is 
\beq
  p_{xy} = C_N \int_0^1 {\rm d} r_1 \cdots  \int_0^1 {\rm d} r_N \: r_x^2 \, r_y^2 \,
  \delta(1- {\bf r}^2) \, ,
  \label{eq:int}
\eeq
where ${\bf r}^2 = r_1^2 + \cdots + r_N^2$, and
$C_N^{-1} \equiv \int_0^1 {\rm d} r_1 \cdots  \int_0^1 {\rm d} r_N \, \delta(1- {\bf r}^2)$
is a normalization constant.
Changing the limits of integration due to the delta function and
using the property of the Gamma function $\Gamma(z) = \int_0^{+ \infty} y^{z-1} e^{-y} {\rm d}y$, 
it is easy to show that $C_N = 2^N \pi^{-N/2} \Gamma (N/2)$ and
\beq
  p_{xy} = \frac{1+2\, \delta_{x,y}}{N(N+2)} \, .
  \label{eq:probab}
\eeq

\section{}\label{appB}
For times longer than those in the scales of Figs.~\ref{fig:L50_Bvar} 
and~\ref{fig:B100_Nvar}, time revivals of the fidelity are present, i.e. the 
fidelity increases back towards the unit value periodically, due to the finite
system size. 
For finite values of the transverse field revivals are not perfect, 
that is $\fidav(t) \ne 1$ for $t > 0$. 
Anyway, as far as $\lambda$ increases, the revivals are stronger
and happen with period $t^R$ which does not depends on the system size $n$.
This can be understood in the limit $\lambda \to +\infty$, where the ground
state of the environment $\ket{\varphi}_E$ is a fully $z$-polarized state,
thus being an eigenstate of the chain Hamiltonian $\Ham^{x}_E$
in Eq.~(\ref{eq:HamChain}): the generalized Loschmidt echo 
of Eq.~(\ref{eq:GenEcho}) is then given by $L_{xy} = e^{- i \eps \mathcal{N}_{x-y} t}$,
where $\mathcal{N}_{x-y}$ is the number of excited qubits in the sequence
$x$ minus the one in the sequence $y$.
It is easy to see that there are $2^n \times {n \choose k}$ different possibilities 
to choose two sequences $x, y$ such that the corresponding states 
differ in the state of $k$ qubits, then
$\mathcal{N}_{x-y} = \pm j$ with $j = 0,2, \ldots , k$
(if $k$ is even) or $j =1,3, \ldots , k$ (if $k$ is odd).
Therefore, when averaging over input states, each term contributes
with $p_{xy} \, e^{\pm i \, \eps j t}$.
Noting that $( {\cal U}_y^\dagger \, {\cal U}_x )^\dagger =
{\cal U}_x^\dagger \, {\cal U}_y $, we have
\beq
\langle F \rangle = 2^n p_{xx} +
p_{xy} \left( c_0 + c_1 \cos (\eps t) + c_2 \cos (2 \eps t)
+\ldots + c_n \cos (n \eps t) \right)\;,
\eeq
It follows a perfect revival for the fidelity at times $t^R$ such that
$\eps t^R = 2 \pi \; [ {\rm mod} \;2 \pi ].$

\end{document}